\RequirePackage{fix-cm}
\documentclass[smallextended]{svjour3}       % onecolumn (second format)
\smartqed  % flush right qed marks, e.g. at end of proof
\usepackage{graphicx}
 \journalname{JSP}

\usepackage[colorlinks]{hyperref}
\usepackage{empheq}
\usepackage[scr]{rsfso}
\usepackage[numbers,sort&compress]{natbib}

\newcommand{\sref}[1]{Section \ref{#1}}
\newcommand{\fref}[1]{Figure \ref{#1}}

\DeclareMathOperator\arcsinh{arcsinh}

\begin{document}

\title{Large deviations in the symmetric simple exclusion process with slow boundaries. \thanks{This work is dedicated to Joel Lebowitz on the occasion of his 90th birthday. By his scientific achievements, his influence, and his committments in the defense of human rights, Joel is and will remain an incomparable example for generations of scientists. He is also a very dear friend to many of us. }}

\author{Bernard Derrida \and Ori Hirschberg \and Tridib Sadhu}

\institute{Bernard Derrida \at
              Coll\`{e}ge de France, 11 place Marcelin Berthelot, 75005 Paris, France.\\
              Laboratoire de Physique de l'\'{E}cole Normale Sup\'erieure, ENS, \\
              Universit\'{e} PSL, CNRS, Sorbonne Universit\'{e}, Universit\'{e} de Paris, F-75005 Paris, France.\\
              \email{bernard.derrida@college-de-france.fr}
              \and
              Ori Hirschberg \at
              Courant Institute of Mathematical Sciences, New York University, \\
              New York, New York 10012, USA.\\
              \email{ori.hirschberg@beyondminds.ai}
           \and
          Tridib Sadhu  \at
          Department of Theoretical Physics, Tata Institute of Fundamental Research, \\ Homi Bhabha Road, Mumbai 400005, India.\\
          \email{tridib@theory.tifr.res.in}
}
\date{Received: date / Accepted: date}

\maketitle

\begin{abstract}
We obtain the exact large deviation functions of the density profile and of the current, in the non-equilibrium steady state of a one dimensional symmetric simple exclusion process coupled to boundary reservoirs with slow rates. Compared to earlier results, where rates at the boundaries are comparable to the bulk ones, we show how macroscopic fluctuations are modified when the boundary rates are slower by an order of inverse of the system length.

\end{abstract}

%\PACS{05.40.-a, 05.70.Ln, 05.10.Gg}
%\date{\today}
\maketitle

\section{Introduction}

Over the past few decades, studies of large deviations have attracted a lot of interest in the theory of non-equilibrium systems \cite{Derrida2007,Touchette2009,Bertini2014,Mallick2015}. Large deviation functions (ldf) or rate functions, as they are often called, characterize macroscopic fluctuations for an extended system in a similar way that the free energy does in thermodynamics. For example, generic long-range correlations outside equilibrium can be understood in terms of non-local ldf \cite{Bertini2007Correlation,Derrida2007,Bodineau2008,Bertini2009}, phase transitions are associated with singularities of ldf \cite{Bodineau2005,Bunin2012,Bunin2013,TsobgniNyawo2016a}, and fluctuation relations of ldf \cite{Evans1993,Gallavotti1995,Jarzynski1997,Hurtado2011} extend the fluctuation-dissipation theorem far beyond the linear response regime. This way ldf offer a possible extension of thermodynamic potentials outside equilibrium.

For extended systems, it is still difficult to determine the ldf, even in numerical simulations \cite{Giardina2006,Lecomte2007a,Gorissen2012b,Hurtado2014,Bunin}. There are however few cases of interacting particles for which analytical results of ldf are available. Some well-known examples are exclusion processes in one dimension \cite{Derrida2001,Derrida2002a,Bertini2002,Tailleur2007,Bahadoran2010,Carinci2013,Derrida2004,Bodineau2004,Bodineau2007,Bertini2005,
Bertini2006Current,Enaud2004,Prolhac2008,Prolhac2009,Gorissen2012,Lazarescu2013,Lecomte} and their higher dimensional extensions \cite{Akkermans2013}, the Kipnis Marchioro Presutti model of heat conduction \cite{Kipnis1982,Bertini2005KMP}, and the zero-range process \cite{Harris2005}. Besides microscopic solutions, these examples can be solved using hydrodynamic approaches \cite{Bertini2001,Bertini2002,Bertini2003SEP,Bertini2014,DerridaMFT,Bertini2007,Tailleur2007} which potentially could be generalized to a wider class of systems.  

In the analogy of ldf as a thermodynamic potential it is important to understand the effect of boundaries.  In equilibrium with short-range interactions, the bulk free energy does not depend on precise details of the coupling to the reservoir. For transport properties, like current fluctuations, they do depend. Outside equilibrium, even the steady state fluctuations are sensitive to the boundary conditions. A natural question is how sensitive or robust are the fluctuations, particularly the associated ldf, to the details at the boundary.

Our main interest in this paper is to quantify the dependence of ldf against changes in the boundary. For this we consider a well-known model, the symmetric simple exclusion process (SSEP) with \emph{slow} coupling with two reservoirs at the boundaries and determine the ldf of the density and of  the current, in the non-equilibrium steady state. Our results for the density are derived from exact expressions of the steady state correlations functions \cite{Derrida20072}. For the current, the ldf is obtained using an additivity principle \cite{Bodineau2004,Bodineau2007}, which we have verified using an exact low density expansion \cite{Derrida2019} up to fourth order.

\begin{figure}
 \centering \includegraphics[width=0.75\textwidth]{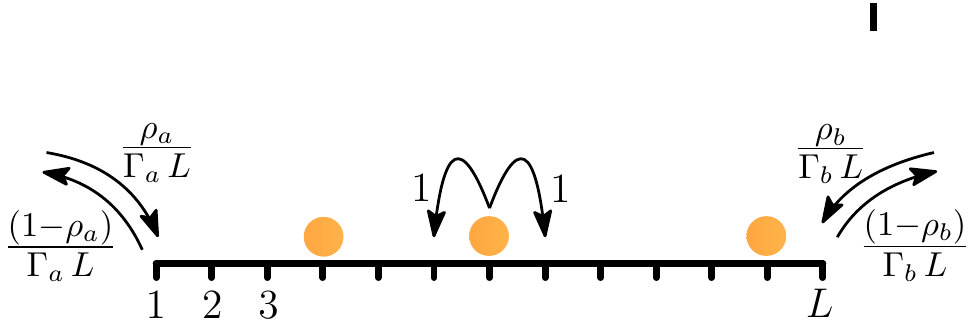}
  \caption{A one-dimensional symmetric simple exclusion process of $L$ sites coupled at the left and the right boundaries with reservoirs at densities $\rho_a$ and $\rho_b$, respectively. Here, \emph{slow} boundaries means that input and exit rates at the boundaries are of order $\frac{1}{L}$.\label{fig:SEP}}
\end{figure}

Our results show that \emph{slow} boundaries as in \fref{fig:SEP}, \textit{i.e.} when the rates at the boundaries are of the order of the inverse of the system length $L$, is the marginal case. For boundary rates faster than $\mathcal{O}(L^{-1})$ bulk fluctuations are given by the ldf for fast coupling, where they are independent of details at the boundary. For slower rates the system is effectively in equilibrium.

The SSEP has been extensively studied in the physics \cite{Derrida2007,Bertini2014,Mallick2015} and in the mathematics \cite{Liggett1999,DeMasi1982a,Landim2006,Bernardin2010} literature, as well as a model for biological transport \cite{Chou1998}. For slow coupling, the earliest study in \cite{Baldasso2017} established that the average macroscopic density profile in the steady state is a solution of the heat equation with Dirichlet, Robin, or Neumann boundary conditions, depending on the speed of the rates at the boundary. The work was later extended in \cite{Franco2019,Franco2016,Goncalves2018,Tsunoda2019}, particularly for small fluctuations in the steady state. Variants of the model with slow rates have also been studied in \cite{Bodineau2009,DeMasi2011,DeMasi2012,Franco2013,Redig2011,Landim2018,Erignoux2019}.  

The rest of this paper is organized as follows. After a description of the microscopic dynamics of the model, we recall earlier results of ldf for the density and for the current in the fast coupling regime. We then present our new results for the ldf in the slow coupling regime. In the subsequent sections, we show how these results are obtained, first for the density, and then for the current.

\section{System}
We consider a SSEP on a finite one-dimensional lattice of $L$ sites. Inside the lattice, particles hop between nearest neighbor sites with unit jump rates following simple exclusion such that a site is occupied by at most one particle, at a given instance. The occupation of a site $i$ at time $\tau$ is denoted by $n_i(\tau)$, which is $1$ if the site is occupied and $0$ if it is empty. The system is coupled with particle reservoirs at density $\rho_a$ and $\rho_b$ at the left and  the right boundaries. This is modeled by deposition and evaporation of particles at the boundary sites $i=1$ and $i=L$, with rates shown in \fref{fig:SEP}.

\section{Earlier results for fast coupling}
Ldf for this model have been determined \cite{Derrida2001,Derrida2002a,Bertini2002,Tailleur2007,Derrida2004,Bodineau2004,Bodineau2007,Bertini2005,Bertini2006Current} in the fast coupling regime, where jump rates at the boundaries are comparable to the bulk. Precisely, it corresponds to the case $\Gamma_a=\frac{\gamma_a}{L}$ and $\Gamma_b=\frac{\gamma_b}{L}$ where $\gamma_a$ and $\gamma_b$ are $\mathcal{O}(1)$ for large $L$. It was shown \cite{Derrida2001,Derrida2002a,Bertini2002,Tailleur2007} that in the hydrodynamic limit, with $x= \frac{i}{L}$ for large $L$, the steady state probability of a macroscopic density profile $ \rho(x)$ has a large deviation description,
\begin{equation}
P[\rho(x)]\sim e^{-L \; \psi[\rho(x)]}, \label{eq:ldp density}
\end{equation}
where $\psi[\rho(x)]$ is the ldf of density. Here, the symbol $\sim$ means that the ratio of logarithm of two sides in \eqref{eq:ldp density} converges to $1$ for $L\to \infty$.  (By macroscopic density profile $\rho(x)$, we mean that if a large system of size $L$ is decomposed into a large number $\frac{L}{\ell}$ of large boxes of size $\ell$, then $P[\rho(x)]$ is the sum of the weights of all the configurations with $\lfloor \ell \, \rho(x)\rfloor $ particles in the box near position $x$.)

\begin{subequations}
For fast coupling, $\psi[\rho(x)]$ has a variational expression \cite{Derrida2001,Derrida2002a,Bertini2002,Tailleur2007,Bertini2014},
\begin{equation}
\psi[\rho(x)]= \psi_\text{fast}[\rho(x)]\equiv\max_{F(x)} \int_0^1dx\;B\big(\rho(x),F(x)\big),\label{eq:psi str} 
\end{equation}
where
\begin{eqnarray}
B\big(\rho(x),F(x)\big)=&&\rho(x)\ln \frac{\rho(x)}{F(x)}+ \nonumber \\
&&\big(1-\rho(x)\big)\ln \frac{1-\rho(x)}{1-F(x)} +\ln \frac{F'(x)}{\rho_b-\rho_a},
\end{eqnarray}
and the maximum is over a differentiable monotone function $F(x)$ for $0\le x\le 1$ with boundary conditions $F(0)=\rho_a$ and $F(1)=\rho_b$. Note that the ldf \eqref{eq:ldf for SSEP strong} does not depend on $\gamma_a$ and $\gamma_b$.
\label{eq:ldf for SSEP strong}
\end{subequations}

The time integrated particle current $\mathcal{Q}_\mathcal{T}$ measured over a time duration $\mathcal{T}$ has an analogous large deviation description \cite{Derrida2004,Bodineau2004,Bodineau2007,Bertini2005,Bertini2006Current,Lecomte}. For large $\mathcal{T}$ and large $L$ with $\mathcal{T}\gg L^2$, the probability has the form
\begin{equation}
P\left(\frac{\mathcal{Q}_\mathcal{T}}{\mathcal{T}}=j \right)\sim e^{-\frac{\mathcal{T}}{L} \phi(L\, j)}, \label{eq:PL Q fast ldf}
\end{equation}
with $\phi(q)$ being the ldf of current. It has a simple expression when written as a Legendre transformation
\begin{equation}
\phi(q)=\max_\lambda \left\{ q\; \lambda -\mu(\lambda)\right\} \label{eq:Legendre}
\end{equation}
where $\mu(\lambda)$ is the scaled cumulant generating function of current, such that, 
\begin{equation}
\langle e^{\lambda\mathcal{Q}_\mathcal{T}}\rangle \sim e^{\frac{\mathcal{T}}{ L}\mu(\lambda)} \label{eq:ldf mu}
\end{equation}
for $\mathcal{T}\gg L^2\gg 1$. Similar to \eqref{eq:ldf for SSEP strong}, $\mu(\lambda)$ does not depend on $\gamma_a$ and $\gamma_b$. Rather, surprisingly, it  depends on $\lambda$, $\rho_a$ and $\rho_b$ through a single parameter 
\begin{equation}
\omega(\lambda,\rho_a,\rho_b)=\rho_a(1-\rho_b)(e^\lambda-1)+\rho_b(1-\rho_a)\left(e^{-\lambda}-1\right).\label{eq:omega}
\end{equation}
For fast coupling, it was shown \cite{Derrida2004,Bodineau2004,Bodineau2007,Bertini2005,Bertini2006Current,Lecomte} that 
\begin{subequations}\label{eq:current ldf strong}
\begin{eqnarray}
\mu(\lambda)&=&R_\text{fast}\left(\omega\left(\lambda,\rho_a,\rho_b\right)\right) \quad \text{with} \label{eq:mu R} \\
R_\text{fast}(\omega)&=&\left(\arcsinh\sqrt{\omega}\right)^2. \label{eq:R str}
\end{eqnarray}
\end{subequations}

\section{New results for slow coupling}
Our new results in the present paper are for slow boundaries which correspond to $\Gamma_a\sim \mathcal{O}(1)$ and $\Gamma_b\sim \mathcal{O}(1)$ for large $L$.
We show that the probability of a macroscopic density profile $\rho(x)$ has a similar large deviation form \eqref{eq:ldp density} with \eqref{eq:ldf for SSEP strong} replaced by
\begin{eqnarray}
\psi[\rho(x)]=\psi_\text{slow}[\rho(x)]&&\equiv \max_{F(x)}\bigg\{\int_0^1 \! \!\! dx\; B\left(\rho(x),F(x)\right)\nonumber \\
&&\left.+\Gamma_a \ln \frac{F(0)-\rho_a}{\Gamma_a(\rho_b-\rho_a)}+\Gamma_b \ln \frac{\rho_b-F(1)}{\Gamma_b(\rho_b-\rho_a)}\right\}, \label{eq:ldf for SSEP}
\end{eqnarray}
where the maximization is over differentiable monotone function $F(x)$ for $0\le x\le 1$.

\begin{subequations}\label{eq:equation for F full}
The maximization gives $F(x)$ as a monotone solution of
\begin{equation}
\rho(x)=F(x)+\frac{F(x)(1-F(x))F''(x)}{F'(x)^2}, \label{eq:rho F relation}
\end{equation}
for $0\le  x \le 1$ with a Robin boundary condition 
\begin{equation}
F(0)=\rho_a+\Gamma_a\, F'(0)\quad \text{and}\quad F(1)=\rho_b-\Gamma_b\, F'(1).\label{eq:continuity of F}
\end{equation}
\end{subequations}

Note that unlike in \eqref{eq:psi str}, for the variational formula \eqref{eq:ldf for SSEP}, $F(x)$ is no longer fixed at the boundaries. The condition \eqref{eq:continuity of F} comes from the maximization in \eqref{eq:ldf for SSEP}.

It is straightforward to see that the optimal density profile $\bar{\rho}(x)$, \textit{i.e.} the profile which minimizes \eqref{eq:ldf for SSEP} satisfies $\bar{\rho}(x)=F(x)$, so that
\begin{equation}
\bar \rho (x)=\rho_a \left(1-\frac{x+\Gamma_a}{1+\Gamma_a+\Gamma_b} \right)+\rho_b \frac{x+\Gamma_a}{1+\Gamma_a+\Gamma_b}, \label{eq:rho bar weak}
\end{equation}
in agreement with the solution of the hydrostatic equation derived in \cite{Baldasso2017}.

The probability of the current also has a large deviation description \eqref{eq:PL Q fast ldf} with $\mu(\lambda)$ in \eqref{eq:Legendre} given by
\begin{subequations}\label{eq:mu slow full}
\begin{equation}
\mu(\lambda)=R_{\text slow}\left(\omega\left(\lambda,\rho_a,\rho_b\right)\right), \label{eq:mu slow}
\end{equation}
where $R_{\text slow}(\omega)$ can be written in a variational formula
\begin{equation}
R_{\text slow}(\omega)=\min_{t_a,t_b}\left\{\frac{\sinh^2 t_a}{\Gamma_a}+(u-t_a-t_b)^2+\frac{\sinh^2 t_b}{\Gamma_b}\right\} \label{eq:R variational}
\end{equation}
with $ \omega=\sinh^2u$ and $\omega$ given by \eqref{eq:omega}. 
\end{subequations}

For both ldf, the fast coupling results \eqref{eq:ldf for SSEP strong} and \eqref{eq:current ldf strong} can be recovered as the $\Gamma_{a(b)}\to 0$ limit of \eqref{eq:ldf for SSEP} and \eqref{eq:mu slow full}. One way to see this is by solving the optimization problem in (\ref{eq:ldf for SSEP},\,\ref{eq:mu slow full}) by a perturbation expansion in powers of small $\Gamma_{a(b)}$. In the rest of this paper, we show how \eqref{eq:ldf for SSEP} and \eqref{eq:mu slow full} are obtained, in this order. 

\section{Derivation of (\ref{eq:ldf for SSEP}) for the density \label{sec:ldf for SSEP}}
For the symmetric simple exclusion process the probability distribution of the occupation variables $\{n_i\}$ in the steady state is known in terms of the matrix product ansatz \cite{Derrida2001,Derrida2002a,Derrida2007,Blythe2007}. It has been shown (Eq.(A7) in \cite{Derrida20072}) that all the correlations of $n_i$ satisfy the following recursion relations on the system size: correlation functions $\langle n_{i_1}\cdots n_{i_{k}}\rangle_{_{L}}$, for a system of length $L$ with rates given as in \fref{fig:SEP}, can be expressed in terms of lower correlation functions $\langle n_{i_1}\cdots n_{i_{k-1}}\rangle_{_{L}}$ for the same system and $\langle n_{i_1}\cdots n_{i_{k-1}}\rangle_{_{L-1}}$ for a system of length $L-1$ keeping the rates at the boundaries unchanged.
\begin{eqnarray}
\langle n_{i_1} \cdots n_{i_{k}}\rangle_{_L}=&&\rho_b\, \langle n_{i_1}\cdots n_{i_{k-1}}\rangle_{_{L}}+\nonumber \\
&&\frac{(\rho_a-\rho_b)}{N}\left(L+\Gamma_b L-i_{k}\right)\langle n_{i_1}\cdots n_{i_{k-1}}\rangle_{_{L-1}},
\label{eq:micro corr}
\end{eqnarray}
where we denote 
\begin{equation}
N=L-1+\Gamma_a L+\Gamma_b L.
\end{equation}
This relation allows to determine all the correlations for the system with slow boundaries. For example,
\begin{equation}
\langle n_i \rangle_{_L}=\rho_a \left(\frac{L+\Gamma_b\, L-i}{N} \right)+\rho_b \left(\frac{i+\Gamma_a\, L-1}{N}\right),
\end{equation}
which gives back \eqref{eq:rho bar weak} in the hydrodynamic limit $x=\frac{i}{L}$, for large $L$. Similarly for $i<j$, the connected correlation function,
\begin{equation}
\langle n_i n_j \rangle_{_L} - \langle n_i \rangle_{_L}\langle n_j \rangle_{_L}= -\frac{(\rho_a-\rho_b)^2}{N-1}\left(\frac{i+\Gamma_a\, L-1}{N}\right)\left( \frac{L+\Gamma_b\, L-j}{N}\right)
\end{equation}
generalizes the known expression \cite{Spohn1983} to the slow boundary case.

\begin{figure}
 \centering \includegraphics[width=0.75\textwidth]{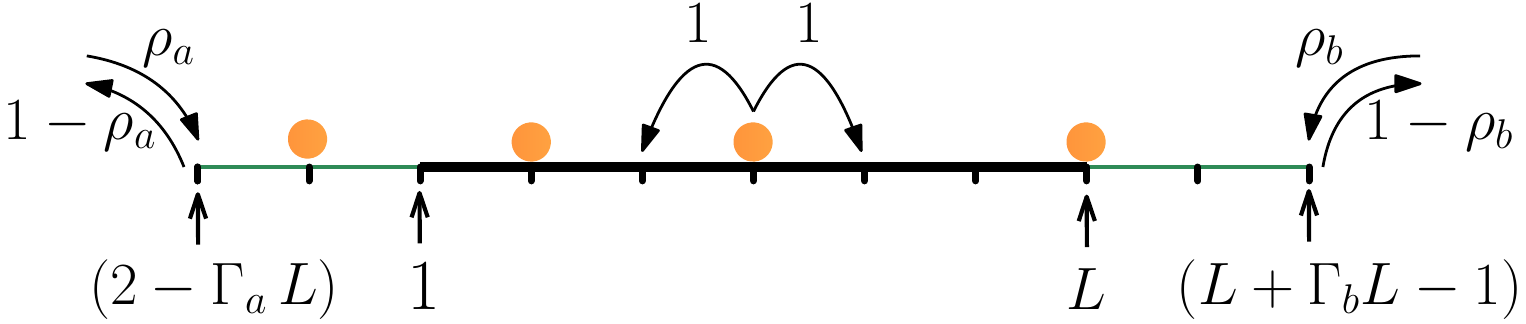}
  \caption{A symmetric simple exclusion process on an `extended' chain of length $ L-2+\Gamma_a L+\Gamma_b L$ indexed by $i\equiv \{2-\Gamma_a L,\ldots, L+\Gamma_b L-1\}$. Here, the jump rates at the boundaries represent fast coupling with reservoirs of density $\rho_a$ and $\rho_b$ at the left and right boundary, respectively. The central part (heavy line) of this extended system with fast boundaries has the same ldf of the density as the system of \fref{fig:SEP} with slow boundaries. \label{fig:extended micro system}}
\end{figure}

Based on relation \eqref{eq:micro corr} one can see that, for sites $1\le i\le L$,
\begin{equation}
\langle n_{i_1} \ldots n_{i_k}\rangle_{_{L}}=\langle n_{i_1} \ldots n_{i_k}\rangle_{_{N-1}}^\text{(fast)} ,
\label{eq:micro cum reln}
\end{equation}
where the right hand side is the correlation function of the occupation variables of an extended system of length $N-1$ with the leftmost site at position $i=2-\Gamma_a \,L$ and the rightmost site  at position $i=L+\Gamma_b\, L-1$. These two boundary sites are coupled to the reservoirs by fast rates with $\Gamma_{a(b)}=\frac{1}{L}$ shown in \fref{fig:extended micro system}. Therefore, a system of size $L$ with slow boundaries is equivalent to the central part (of $L$ sites) of a larger system of size $N-1$ with fast boundaries.

The relation \eqref{eq:micro cum reln} leads to a similar relation for the probability of the occupation variables. 
\begin{equation}
P_L(n_1,\cdots,n_L)=\sum_{\{n_i\}}^\prime P_{N-1}^\textrm{(fast)}(n_{2-\Gamma_a L},\cdots,n_{L+\Gamma_b L-1}),
\label{eq:Prob micro relation}
\end{equation}
where $P_{N-1}^\textrm{(fast)}(\{n_i\})$ is the steady state probability for the strongly coupled extended system in \fref{fig:extended micro system}, and the prime in $\sum^\prime$ denotes summation over $n_i=\{0,1\}$ from sites in the extended parts, $i=\{2-\Gamma_a L,\cdots,0\}$ and $i=\{L+1,\cdots,L+\Gamma_b L-1\}$.

The large deviation form in (\ref{eq:ldp density},\,\ref{eq:ldf for SSEP strong}) implies that $P_{N-1}^\textrm{(fast)}$ for the extended system has similar asymptotics. In the hydrodynamic scale $x=\frac{i}{L}$ for large $L$, the probability of a density profile $r(x)$ is
\begin{equation}
P_{N-1}^{\textrm {(fast)}}[r(x)]\sim e^{-L\, \Psi_\textrm{fast}[r(x)]} \label{eq:ldf str}
\end{equation}
where $x$ ranges from $-\Gamma_a$ to $1+\Gamma_b$. The large deviation function $\Psi_\text{fast}$ for this extended geometry is given in \eqref{eq:ldf for SSEP strong},
\begin{equation}
\Psi_\text{fast}[r(x)]=\max_{F(x)} \int_{-\Gamma_a}^{1+\Gamma_b}dx\;B(r(x),F(x))\label{eq:psi str extended}
\end{equation}
with the boundary conditions $F(-\Gamma_a)=\rho_a$ and $F(1+\Gamma_b)=\rho_b$.  

\begin{figure}
 \centering \includegraphics[width=0.75\textwidth]{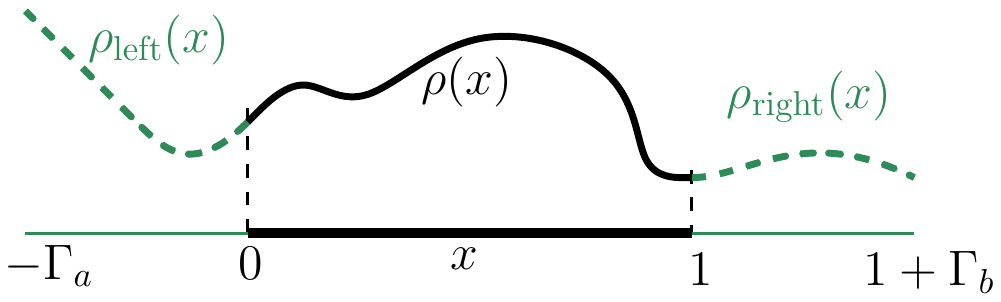}
  \caption{Schematics of a macroscopic density profile $r(x)$ in \eqref{eq:density partition} for an extended SSEP with fast boundaries of density $\rho_a$ at $x=-\Gamma_a$ and of density $\rho_b$ at $x=1+\Gamma_b$. \label{fig:marginalProfile}}
\end{figure}

Therefore, \eqref{eq:Prob micro relation} together with the asymptotics \eqref{eq:ldp density} and \eqref{eq:ldf str} gives
\begin{subequations}
\begin{equation}
\psi[\rho(x)]=\min_{\rho_{\text{left}}(x)}\min_{\rho_{\text{right}}(x)} \Psi_\text{fast}[r(x)],
\label{eq:marginal ldf}
\end{equation}
where we denote
\begin{equation}
r(x)=\begin{cases}\rho_\text{left}(x) \qquad & \textrm{for $-\Gamma_a \le x < 0$,}\cr
\rho(x) \qquad & \textrm{for $0 \le  x \le  1$,}\cr
\rho_\text{right}(x) \qquad &\textrm{for $1 < x \le 1+\Gamma_b$.}\end{cases} \label{eq:density partition}
\end{equation}
The corresponding geometry is sketched in \fref{fig:marginalProfile}. 
\label{eq:main result}
\end{subequations}

When evaluating \eqref{eq:marginal ldf} the optimization over $(\rho_\text{left},\rho_\text{right})$ and $F(x)$ gives for the optimal density in the extended parts,
\begin{subequations}\label{eq:rho eq F}
\begin{eqnarray}
\rho_\text{left}(x)=F(x) \quad & \text{for $x< 0$,}\\
\rho_\text{right}(x)=F(x) \quad & \text{for $x> 1$.}
\end{eqnarray}\label{eq:extended rho F}\end{subequations}
The optimal $F(x)$ then satisfies
\begin{equation}
\frac{F''(x)}{F'(x)^2}=\begin{cases}\frac{\rho(x)-F(x)}{F(x)(1-F(x))} \quad & \textrm{for $0 \le x \le 1$,} \\
0 \quad & \textrm{for extended parts.}\end{cases}\label{eq:Optimal F}
\end{equation}
This means that, in the extended parts, $F(x)$ is linear,
\begin{equation}
F(x)=\begin{cases}F(0)+(F(0)-\rho_a)\; \frac{x}{\Gamma_a} \quad & \textrm{for $x < 0$,} \\
F(1)+(\rho_b-F(1))\;\frac{(x-1) }{\Gamma_b} \quad & \textrm{for $x>1$,}\end{cases} \label{eq:Explicit F outside}
\end{equation}
where we used the fixed boundary condition for fast couplings, $F(-\Gamma_a)=\rho_a$ and $F(1+\Gamma_b)=\rho_b$. 

\begin{figure}
  \centering \includegraphics[width=0.75\textwidth]{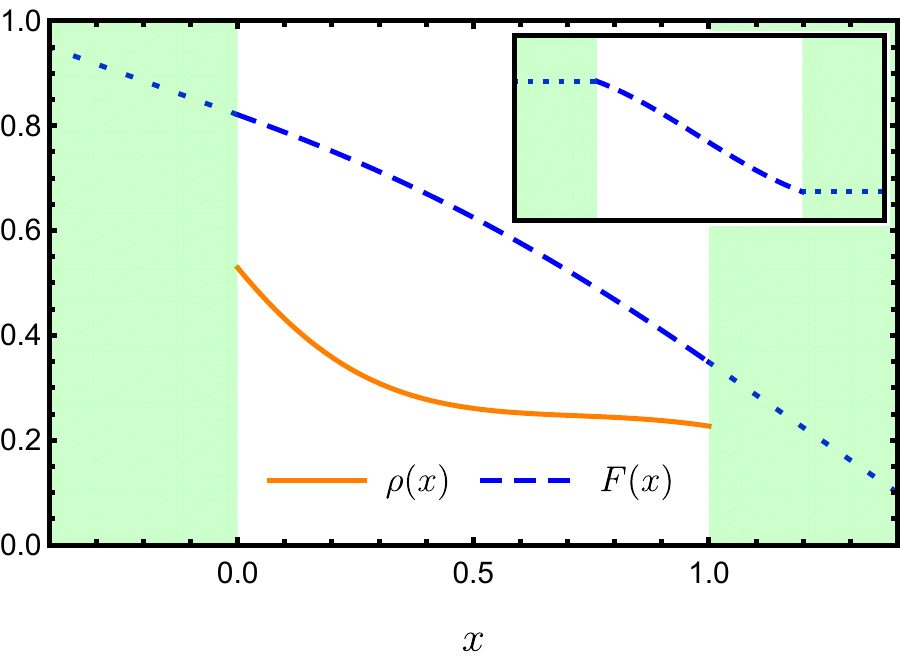}
  \caption{Optimal $F(x)$ for \eqref{eq:ldf for SSEP} corresponding to a density profile $\rho(x)$ in $0\le x\le 1$ with reservoirs of density $\rho_a=0.95$ and $\rho_b=0.1$ with parameters $\Gamma_a=\Gamma_b=0.4$. The shaded regions outside $0\le x\le 1$ indicate extended parts for the construction in \eqref{eq:main result}. The dotted line in the shaded regions indicates the linear part of $F(x)$ in \eqref{eq:Explicit F outside}, which coincide with the optimal density \eqref{eq:extended rho F}.  The inset shows the first derivative of $F(x)$.  \label{fig:rhoForSEPa}}
\end{figure}

Then, the continuity of $F(x)$ and $F'(x)$ at $x=0$ and $x=1$ gives the condition \eqref{eq:continuity of F}, whereas substituting (\ref{eq:rho eq F},\,\ref{eq:Explicit F outside}) in \eqref{eq:main result} gives the expression in \eqref{eq:ldf for SSEP}. The optimal $F(x)$ for a sample density profile $\rho(x)$ is shown in \fref{fig:rhoForSEPa}.

\section{Derivation of (\ref{eq:mu slow full}) for the current}
It is well known that, for a system of size $L$, $\mu(\lambda)$ defined in \eqref{eq:ldf mu} can be obtained \cite{Derrida2004,Derrida2007,Derrida2019,Prolhac2008,Prolhac2009,Prolhac2010,Gorissen2012,Lazarescu2013,Lazarescu2015} from the largest eigenvalue $\Lambda_L(\lambda)$ of a tilted matrix. For a diffusive system like the SSEP, one expects \cite{Derrida2007,Bertini2014} that $\Lambda_L(\lambda)\simeq \frac{1}{L}\mu(\lambda)$, for large $L$. For an \textit{asymmetric} simple exclusion process with an arbitrary set of rates at the boundary, $\Lambda_L(\lambda)$ has been obtained exactly in \cite{Gorissen2012,Lazarescu2013,Lazarescu2015}. The solution for $\Lambda_L(\lambda)$ is written in a parametric form and in terms of a functional equation, which is not easy to solve. Taking then the limit of the SSEP and the large $L$ limit are additional difficulties hard to overcome.

Here, we obtain directly $\mu(\lambda)$ in \eqref{eq:mu slow full} following an additivity principle \cite{Bodineau2004,Bodineau2007}. For fast couplings, the additivity principle is known to give \cite{Bodineau2004,Bodineau2007,Bertini2005,Bertini2006Current} the correct result \eqref{eq:current ldf strong} and its validity has been mathematically justified \cite{Bertini2005,Bertini2006Current} as well as numerically confirmed in other models \cite{Hurtado2009,Hurtado2010,Hurtado2014}. 

For slow boundaries, the additivity principle can be stated as follows. Consider the system in \fref{fig:SEP} as composed of three subsystems: the \textit{bulk}, which consists of all bonds between sites $i=1$ to $i=L$, and the two bonds at the \textit{left} and at the \textit{right} boundaries that link the boundary sites to the reservoirs. We shall call these subsystems $S_\text{bulk}$, $S_\text{left}$, and $S_\text{right}$. These are linked to each other by the sites $i=1$ and $i=L$. The main idea of the additivity principle is to assume that, for large $L$, the correlations between these three subsystems can be neglected. This way, the probability \eqref{eq:PL Q fast ldf} of integrated current $\mathcal{Q}_\mathcal{T}\simeq j\, \mathcal{T}$ for $\mathcal{T}\gg L^2\gg 1$, can be written as a product of its probability in the subsystems
\begin{equation}
P(j)\simeq  \max_{\rho_0,\rho_1} P_\textrm{left}(j,\rho_a,\rho_0)P_{\textrm{bulk}}(j,\rho_0,\rho_1)   P_{\textrm{right}}(j,\rho_1,\rho_b) , \label{eq:variational prob subsystem}
\end{equation}
where $P_\text{S}(j,r,s)$ is the probability of integrated current $\mathcal{Q}_\mathcal{T}\simeq j\, \mathcal{T}$ in the subsystem $\text{S}$ with densities $r$ and $s$ at its left and right boundaries; in \eqref{eq:variational prob subsystem} one considers that the three subsystems are independent except that $\rho_0$ and $\rho_1$ are adjusted to make $P(j)$ maximum \cite{Bodineau2004,Bodineau2007}.

For $\mathcal{T}\gg L^2\gg 1$, the probability $P_{\textrm{bulk}}(j,\rho_0,\rho_1)$ has the asymptotics \eqref{eq:PL Q fast ldf} of a system with fast coupling with reservoirs of density $\rho_0$ and $\rho_1$,
\begin{equation}
P_\textrm{bulk}(j,\rho_0,\rho_1)\sim e^{-\frac{\mathcal{T}}{L} \phi_\text{fast}(j \, L,\rho_0,\rho_1)}\label{eq:P bulk}
\end{equation}
where $\phi_\text{fast}(q,\rho_0,\rho_1)$ is the Legendre transform \eqref{eq:Legendre} (in $q\leftrightarrow \lambda$ variables) of $R_\text{fast}(\omega(\lambda,\rho_0,\rho_1))$ in \eqref{eq:current ldf strong}.

The following argument may be used to justify \eqref{eq:variational prob subsystem}. For $S_\text{left}$, the rates are slower by an order $\frac{1}{L}$ compared to unit rates inside $S_\text{bulk}$. Therefore, for events in $S_\text{left}$, the sites of $S_\text{bulk}$ near the left boundary are effectively at a local equilibrium, and they act as a reservoir at density $\rho_0$.  Then, $P_\textrm{left}(j,\rho_a,\rho_0)$ for large $L$ is the probability of a current across a single bond with a forward jump rate $\frac{\rho_a(1-\rho_0)}{\Gamma_a L}$ and a backward jump rate $\frac{(1-\rho_a)\rho_0}{\Gamma_a L}$, which corresponds to coupling with reservoirs at densities $\rho_a$ and $\rho_0$. It is straightforward to show that the generating function of the integrated current $\mathcal{Q}_\mathcal{T}$ in time $\mathcal{T}$ across this single bond is $\langle e^{\lambda \mathcal{Q}_\mathcal{T}} \rangle = e^{\frac{\mathcal{T}}{\Gamma_a L}\omega(\lambda,\rho_a,\rho_0) }$
with $\omega$ in \eqref{eq:omega}. This means, for large $L$,
\begin{equation}
P_\textrm{left}(j,\rho_a,\rho_0)\sim e^{-\frac{\mathcal{T}}{\Gamma_a L}\phi_\text{bond}(j\,\Gamma_aL,\rho_a,\rho_0)}, \label{eq:P left}
\end{equation}
where the large deviation function $\phi_\text{bond}(q,\rho_a,\rho_0)$ is the Legendre transform \eqref{eq:Legendre}  (in $q\leftrightarrow \lambda$ variables) of $\omega(\lambda,\rho_a,\rho_0)$. A similar reasoning applies for the subsystem $S_\text{right}$.

Using the asymptotics (\ref{eq:P bulk},\,\ref{eq:P left}) for $\mathcal{T} \gg L^2\gg 1$, \eqref{eq:variational prob subsystem} gives the large deviation function of current for the slowly coupled system
\begin{eqnarray}
\phi_\text{slow}(q,\rho_a,\rho_b)=&&\min_{\rho_0,\rho_1}\bigg\{\frac{\phi_\text{bond}(q\,\Gamma_a,\rho_a,\rho_0)}{\Gamma_a}\nonumber \\
&&+\phi_\text{fast}(q,\rho_0,\rho_1)+\frac{\phi_\text{bond}(q\,\Gamma_b,\rho_1,\rho_b)}{\Gamma_b} \bigg\}.\label{eq:phi additivity}
\end{eqnarray}
Taking the Legendre transform of \eqref{eq:phi additivity} gives a formula for $\mu(\lambda)$ in \eqref{eq:mu slow} with
\begin{eqnarray}
R_\text{slow}(\omega(\lambda,\rho_a,\rho_b))=&& \max_{\rho_0,\rho_1}\;\min_{\lambda_0,\lambda_1}\bigg\{\frac{\omega(\lambda_0,\rho_a,\rho_0)}{\Gamma_a}\nonumber \\
 &&+  R_\text{fast}(\omega(\lambda_1-\lambda_0,\rho_0,\rho_1))+ \frac{\omega(\lambda-\lambda_1,\rho_1,\rho_b)}{\Gamma_b}\bigg\}, \label{eq:variational mu slow final}
\end{eqnarray}
and \eqref{eq:R str}. 

To get the formula \eqref{eq:R variational} one can use the following result (see Appendix for a derivation). For arbitrary differentiable monotone functions $A(\omega)$ and $B(\omega)$ such that the following extremum exists,
\begin{eqnarray}
\max_{\rho_0}\min_{\lambda_0}\bigg\{A(\omega(\lambda_0,\rho_a,\rho_0))&+&B(\omega(\lambda-\lambda_0,\rho_0,\rho_b)) \bigg\}\nonumber \\
&=&\min_{t_a}\bigg\{ A(\sinh^2 t_a)+B(\sinh^2 (t_a\pm u))\bigg\}, \label{eq:optimization idenitity}
\end{eqnarray}
where $\omega(\lambda,\rho_a,\rho_b)=\sinh^2 u$.

Applying \eqref{eq:optimization idenitity} twice in \eqref{eq:variational mu slow final} and using that $R_\text{fast}(\omega)=u^2$ (see \eqref{eq:R str}) one gets
\begin{equation}
R_\text{slow}(\omega(\lambda,\rho_a,\rho_b))= \min_{t_a,t_b}\bigg\{\frac{\sinh^2t_a}{\Gamma_a}+(t_a\pm t_b)^2+\frac{\sinh^2(t_b\pm u)}{\Gamma_b}\bigg\}.
\end{equation}
Both $(\pm)$ solutions have the same minimum and the result, by a simple change of variables, is equivalent to the expression \eqref{eq:R variational}.

We have checked the validity of expression \eqref{eq:R variational} using an exact low density expansion of $\Lambda_L(\lambda)$ up to fourth order, obtained by a perturbation solution of the eigenvalue of the tilted Matrix for SSEP using the method outlined in \cite{Derrida2019}. From \eqref{eq:R variational}, one can expand $R_\text{slow}(\omega)$ in powers of $\omega$ and get
\begin{equation}
R_\text{slow}(\omega)=\frac{\omega}{1+\Gamma_a+\Gamma_b}-\frac{\Gamma_a^3+\Gamma_b^3-(1+\Gamma_a+\Gamma_b)^3 }{3(1+\Gamma_a+\Gamma_b)^4}\omega^2+\mathcal{O}(\omega^3).
\end{equation}
In contrast to the ldf of density, the ldf of the current for the slow boundary system is not related to that of a larger system with fast boundary. In fact, one can notice that already at second order in $\omega$,
\begin{equation}
R_\text{slow}(\omega)\ne \frac{R_\text{fast}(\omega)}{1+\Gamma_a+\Gamma_b}.
\end{equation}
This is in particular because the equivalence with a larger fast boundary system discussed in \sref{sec:ldf for SSEP} does not extend to time correlations.

\section{Conclusion}
In the present work, we have obtained the ldf of the density \eqref{eq:ldf for SSEP} and of the current \eqref{eq:mu slow full} in a SSEP with slow boundary rates $\mathcal{O}(L^{-1})$. Earlier results (\ref{eq:ldf for SSEP strong},\,\ref{eq:current ldf strong}) for the fast coupling regime can be recovered as a special limit $\Gamma_{a(b)}\to 0$ of (\ref{eq:ldf for SSEP},\,\ref{eq:mu slow full}). This means that bulk fluctuations in SSEP are robust for a wide range of boundary rates faster than $\mathcal{O}(L^{-1})$, where they are insensitive to the details of the coupling with reservoirs.

Our derivation for the ldf of the density \eqref{eq:ldf for SSEP} is based on a relation \eqref{eq:Prob micro relation} between a system with slow boundaries and a central part of a larger system with fast boundaries. This is very special of the SSEP. However, the steady state is exactly known for other systems like the WASEP with arbitrary boundary conditions \cite{Enaud2004}. Therefore, one should be able to extend our results for the density to other systems with slow boundaries. For the current, we think that the argument given after \eqref{eq:P bulk} to satisfy \eqref{eq:variational prob subsystem} should remain valid for more general diffusive systems with slow boundaries. An open question would be to try to recover the ldf (\ref{eq:ldf for SSEP},\,\ref{eq:mu slow full}) using the macroscopic fluctuation theory \cite{Bertini2014}.

\appendix
\section{A derivation for \eqref{eq:optimization idenitity}}
It is not immediately clear that $R_\text{slow}$ in \eqref{eq:variational mu slow final} depends on $\lambda$, $\rho_a$, and $\rho_b$ through a single parameter $\omega(\lambda,\rho_a,\rho_b)$. This comes as a result of \eqref{eq:optimization idenitity}. The goal of this appendix is to give a derivation of \eqref{eq:optimization idenitity}.

The optimal $(\rho_0,\lambda_0)$ for \eqref{eq:optimization idenitity} is a solution of
\begin{eqnarray}
A'(\omega_1)\frac{d\omega_1}{d\lambda_0}+B'(\omega_2)\frac{d\omega_2}{d\lambda_0}&=&0, \\
A'(\omega_1)\frac{d\omega_1}{d\rho_0}+B'(\omega_2)\frac{d\omega_2}{d\rho_0}&=&0,
\end{eqnarray}
where we denote $\omega_1\equiv \omega(\lambda_0,\rho_a,\rho_0)$ and $\omega_2\equiv \omega(\lambda-\lambda_0,\rho_0,\rho_b)$. We assume that $A(\omega)$ and $B(\omega)$ are differentiable and monotone. Then, ratio of the two equations gives
\begin{equation}
\frac{d\omega_1}{d\lambda_0}\frac{d\omega_2}{d\rho_0}=\frac{d\omega_2}{d\lambda_0}\frac{d\omega_1}{d\rho_0},
\end{equation}
which can be used to get an explicit expression of $\rho_0=f(\lambda_0,\lambda,\rho_a,\rho_b)$ with the function $f$ being independent of $A(\omega)$ and $B(\omega)$. Using the solution for $\rho_0$ in $\omega_1$ and $\omega_2$, and then eliminating the variable $\lambda_0$ one can express $\omega_2$ in terms of $\omega_1$, $\lambda$, $\rho_a$, and $\rho_b$. A relatively straightforward algebra shows a surprising fact that $\omega_2$ depends on $\lambda$, $\rho_a$, and $\rho_b$ through a parameter $\omega(\lambda,\rho_a,\rho_b)$ and
\begin{equation}
\omega_2= \omega_1+\omega+2\omega_1\,\omega\pm 2\sqrt{\omega_1(1+\omega_1)\omega(1+\omega)}
\end{equation}
with $\omega\equiv \omega(\lambda,\rho_a,\rho_b)$. The expression simplifies by a change of variables $\omega_1=\sinh^2t_a$ and $\omega=\sinh^2 u$ leading to
\begin{equation}
\omega_2= \sinh^2(t_a\pm u).
\end{equation}
From this, \eqref{eq:optimization idenitity} follows immediately. 

\bibliographystyle{spphys}

\bibliography{../../../Mendeley/library}

\end{document}